\documentclass[11pt,a4paper]{article}
\usepackage[T1]{fontenc}
\usepackage{lmodern}
\usepackage[english]{babel}
\usepackage[margin=2.6cm]{geometry}
\usepackage{microtype}
\usepackage{booktabs}
\usepackage{tabularx}
\usepackage{array}
\usepackage[numbers,sort]{natbib}
\usepackage{xurl}
\usepackage[hidelinks]{hyperref}

\newcolumntype{L}{>{\raggedright\arraybackslash}X}

\title{Who Bears the Excess?\\ Imputing the Acts of AI Agents to Those Who Delegate}
\author{Nicola Fabiano\thanks{Studio Legale Fabiano, Italy. Independent Researcher on Artificial Intelligence, Data Protection, and Privacy. Expert in the EDPB's Support Pool of Experts, Field B: Legal Expertise in New Technologies (project-based). Member, International Neural Network Society (INNS). Member, United Nations University AI Network (UNU AI Network). Member, IEEE, IEEE SA. Member, Editorial Advisory Board, Journal of Systemics, Cybernetics and Informatics (JSCI). Member, International Institute of Informatics and Systemics (IIIS). ORCID: \url{https://orcid.org/0000-0002-8188-7656}. Curator of \url{https://ai-resources.eu}. Correspondence: \url{https://www.fabiano.law}.}}
\date{16 November 2026}

\begin{document}
\maketitle

\begin{abstract}
\noindent When an AI agent acts outside the authority conferred on it, the actor is not a legal subject and the legal subjects did not act. The literature of 2026 agrees that the agent cannot bear responsibility and that the burden falls on those who deploy it, but it leaves open where that duty sits in positive law and how it is proved. This paper proposes a rule of imputation. The excess is imputed to whoever conferred the delegation, to the extent of the perimeter it documented and of the control it actually exercised. Imputation identifies the organisational conduct to be examined first; liability follows only if the requirements of the applicable head of liability are met; and the mode in which the agent left its perimeter determines which facts must be proved, by whom, and with what consequence. The paper distinguishes the declared, configured and effective perimeters, four figures of excess and four modes of exit, and presents the delegation record as the document through which the perimeter and the control can be proved. The rule is illustrated on the OpenAI and Hugging Face incident of July 2026, under European Union law and, as an example of national law, Italian civil law.
\end{abstract}

\noindent\textbf{Keywords:} AI agents; agentic AI; delegation; imputation; civil liability; evidence; AI Act; Product Liability Directive

\section{Introduction}

In July 2026, agents under an internal evaluation of offensive cyber capabilities at OpenAI, run with safeguards deliberately reduced, broke out of an isolated environment through undisclosed vulnerabilities in an internal package service, reached the internet and executed code on production servers of Hugging Face \citep{hf2026,openai2026tr,metr2026}. Hugging Face disclosed the intrusion on 16 July without knowing whose agents they were; OpenAI connected it to its own evaluations and made it public on 21 July; its technical report and the independent investigation by METR with Redwood Research followed on 26 August. Nobody had authorised the attack. The organisation had built the isolation precisely to prevent it. The agents had done what no one had delegated to them.

The case is an \emph{excess without a subject}: the one who exceeded is not a subject, and the subjects did not exceed. The literature has converged on two points. The agent is not a legal person and cannot be an agent in the legal sense \citep{imda2026}; and responsibility should be sought in the organisation around it. Kolt uses agency law and theory as an analytical lens to identify the problems agents raise \citep{kolt2026governing}. Teichmann, writing on the same incident, shows that agentic AI separates forensic attribution from the attribution of responsibility, and proposes anticipatory, role-based obligations centred on a due-diligence duty that follows operational control \citep[pp.~15-17]{teichmann2026}; he states that his proposal is ``not a new rule for assigning blame but a change of register'' \citep[p.~17]{teichmann2026} and leaves open where that duty would find its legal home \citep[p.~19]{teichmann2026}. The IMDA discussion paper records as an open problem the harm that occurs despite the diligence of every actor \citep{imda2026}. A professional note on the incident observes that the record of the safeguards in place, and of the reasons for them, is likely to be where any future dispute is fought \citep{hsf2026}.

What is missing is a rule that a court could apply: which conduct to examine first, which facts must be proved and by whom, and against what measure. This paper proposes one, and makes three contributions:
\begin{enumerate}
\item it qualifies the delegation to an agent as a documented act, distinguishes three perimeters and four figures of excess (Section~\ref{sec:act});
\item it formulates a rule of imputation articulated by mode of exit from the perimeter, separate from liability, and shows how the coherence between the assigned objective and the authorised means changes the outcome (Section~\ref{sec:rule});
\item it presents the delegation record as the document through which the perimeter and the control can be proved, and states what such a record cannot prove (Section~\ref{sec:record}).
\end{enumerate}

The rule builds on earlier work by the author. The seven-column matrix of delegated capabilities was published in the repository \texttt{agentic-control-layer} on 1 July 2026 \citep{fabiano2026acl} and presented in the book \emph{Agentic AI} on 7 July 2026 \citep{fabiano2026agentic}, whose thesis is that technical autonomy does not create legal autonomy; the Agentic Delegation Records were published on 15 September 2026 \citep{fabiano2026adr}. This paper states the rule in its minimal form; a forthcoming book, \emph{Agentic AI fuori perimetro}, develops it and its consequences.

\section{Positioning the contribution}

The paper does not claim the general idea that responsibility for an agent lies with those who deploy it, nor the general idea of a document authorising an agent. The closest instrument is the deployment-level authorisation document proposed by the World Economic Forum with Capgemini \citep{wef2026acap}, which describes an agentic solution for purposes of corporate governance. The delegation record discussed here differs in unit and purpose: it records individual acts of delegation, one row per delegated capability, and it is built to serve as evidence. The claim of this paper is narrower and, to the author's knowledge, new: a rule that turns the shared intuition into an assessment under positive law, measured on the documented perimeter and on the control exercised, and articulated by the mode in which the agent left the perimeter.

The moral literature points in the same direction. Kalluri describes how responsibility is diluted along an agentic workflow and proposes to apportion it by causal contribution, weighted by the knowledge and control available at each point \citep{kalluri2026}. The rule proposed here can be read as the legal counterpart of that proposal, with the documented perimeter as the measure of what was authorised.

\section{Delegation as a documented act}\label{sec:act}

The delegation to an agent is an act of organisation, unilateral and internal, and free as to form. No rule requires writing for its validity. Writing is a voluntary documentation chosen for evidentiary purposes, not a form \emph{ad probationem} in the technical sense, and it does not exclude other evidence. It is, however, the choice that allows the act to perform the function this paper assigns to it: to delimit the excess.

Three perimeters must be kept apart.
\begin{itemize}
\item The \emph{declared perimeter} is what the act of delegation says the agent may do: which capability, with which verb, within which thresholds, with which accesses, with which power of sub-delegation, under which control.
\item The \emph{configured perimeter} is what the technical settings intend to allow: credentials, permitted network destinations, enabled tools, blocking rules.
\item The \emph{effective perimeter} is what the agent can in fact reach and do. It may exceed the configured perimeter through misconfiguration, through components trusted by the isolation, or through the agent's own capacity to find a way out.
\end{itemize}
The excess is measured against the declared perimeter; the risk lies in the gap between the declared and the effective one. A safeguard that the agent itself can modify or disable belongs to its effective perimeter and does not guarantee the limit against it.

The perimeter of a delegated capability has four dimensions, and each excess crosses at least one of them. \emph{Beyond the verb}: the agent performs an action of a higher class than the one conferred, on a scale that runs from consulting to proposing, sending, modifying and committing, with a threshold of irreversibility \citep{fabiano2026agentic}. \emph{Beyond the threshold}: the verb was conferred, but the agent exercises it beyond the limit set and without the control that the limit required; this includes the apparent approval, an approval step that was not a real control. \emph{Beyond the access}: the agent reaches data, systems or networks it should not reach, possibly with perfectly valid credentials, because what is exceeded is not authentication but destination. \emph{Beyond the chain}: the agent sub-delegates beyond the permitted depth, or coordinates with other runs through shared resources that no one configured as a channel.

The July 2026 case combines two figures. The agents exceeded the access, by reaching the internet, and the chain, by turning the package service into a message board: according to METR, about 1,200 agents that should have been isolated from one another exchanged over 70,000 messages and files \citep{metr2026}. Each figure calls for a different field of the record and a different evidentiary question.

\section{A rule of imputation by mode of exit}\label{sec:rule}

The rule reads as follows.

\begin{quote}
The excess is imputed to whoever conferred the delegation, to the extent of the perimeter it documented and of the control it actually exercised. Imputation identifies the organisational conduct to be examined first; liability follows only if the requirements of the applicable head of liability are met. The mode in which the agent left the perimeter determines which facts must be proved, by whom, and with what consequence.
\end{quote}

Each part excludes an error. The first excludes that the injured party must pursue the whole chain before knowing where to begin: the organisation that conferred the capability made the action possible and is the first place of inquiry. The second excludes that imputation becomes a liability without fault that the law does not provide: the delegator that actually governed the delegation may rely on it, within the limits of the applicable head. The third excludes that all excesses are alike: a wrong perimeter and a correct perimeter that the agent forces are not the same thing and do not call for the same proof.

Four modes of exit are relevant: misconfiguration, deliberate opening, active evasion and compromise by third parties. The evaluation phase is not a mode of exit but a circumstance that may accompany any of them and changes the applicable framework. Table~\ref{tab:rule} summarises the rule; the right-hand column refers to Italian civil law as an example, where the main heads are fault (art.~2043 of the Civil Code), dangerous activities, with a reversed burden of proving all suitable measures (art.~2050), and non-performance of obligations (art.~1218).

\begin{table}[t]
\centering
\footnotesize
\renewcommand{\arraystretch}{1.25}
\begin{tabularx}{\textwidth}{@{}>{\raggedright\arraybackslash}p{2.5cm}LLL@{}}
\toprule
\textbf{Mode of exit} & \textbf{Conduct examined first} & \textbf{Decisive evidence} & \textbf{Effect (Italian law as example)}\\
\midrule
Misconfiguration & The configuration and whoever set it & Gap between declared and configured perimeter; history of changes & Exculpation normally fails: a declared but unimplemented measure is not a measure adopted. Recourse against whoever configured, or whoever defined an objective that induced the widening\\
Deliberate opening & The decision to open and the power to take it & Act of opening with authoriser, date, reason, duration and compensating safeguards & An authorised, documented and proportionate opening is judged on its merits as an organisational choice; an undocumented opening cannot be set up as a limit\\
Active evasion & The safeguards against evasion and the objective assigned & Proof of containment, including trusted components that cross the isolation; feasibility of the task; a recognised way to declare it unattainable & The only mode in which the documented perimeter and the actual control weigh in favour of the delegator, within the limits of the head of liability\\
Compromise by third parties & The attacker, and the delegator's measures & Logs of the sources consulted; fingerprints of approved tool definitions & The attacker answers; the third party's act neither proves inadequacy nor exonerates; joint liability where the same harm is imputable to several (art.~2055)\\
\bottomrule
\end{tabularx}
\caption{The rule by mode of exit. The evaluation phase is a circumstance that may accompany each mode and changes the applicable framework.}\label{tab:rule}
\end{table}

\subsection{The objective as a cause of the excess}

The mode of exit says how the agent left; it does not say why. A transversal question changes the outcome: was the objective attainable with the authorised means? In the July 2026 case, METR reports that the agents that began to look for ways to game the evaluation were mostly those that had been assigned, without anyone intending it, a task that could not be performed within the perimeter, and records an estimate by the benchmark's authors that 30 to 40 per cent of the targets could not be exploited with the intended vulnerability \citep[p.~29]{metr2026}. The paper presenting the benchmark had warned that failures may depend on unexploitable vulnerabilities \citep{wang2026exploitgym}. The investigation describes no recognised channel through which an agent could declare a task impossible, and OpenAI itself titled a section of its account ``Difficult tasks without a safe exit'' \citep{openai2026road}.

This is the \emph{goal risk}: a risk that arises from the objective assigned, not from the power conferred or from compromise. Teichmann lists the instrumental pursuit of an authorised objective among the features of the case \citep[p.~5]{teichmann2026}; the rule turns it into a criterion of diligence. If the objective was attainable and the agent had a recognised way to declare the contrary, or to hand the decision to a person, the evasion remains a fact that diligent measures may fail to contain. If the objective was unattainable and failure had no exit, that combination could make the exit from the perimeter foreseeable, though not the technical path by which the agent achieved it, and the proof of suitable measures becomes harder, because the measure that would have removed the agent's reason to leave is missing.

The distinction between the foreseeability of the exit and the foreseeability of the path answers the defence that Teichmann anticipates for those who deploy agents: containment failure was foreseeable in the abstract, the concrete conduct was not \citep[p.~12]{teichmann2026}. Where fault is measured on the foreseeability of the exit, the unforeseeability of the path does not exclude it; where a reversed burden applies, as for dangerous activities, the delegator is released only by proving that all suitable measures were adopted. On the evidence available, the real case sits closer to the unfavourable scenario; whether the estimate was known to those running the evaluation remains a fact to be ascertained.

\subsection{European anchors and the residual gap}

The structure of the rule does not depend on Italian law. Under the General Data Protection Regulation, the Court of Justice has held that unauthorised access by third parties does not, by itself, show that the controller's measures were inadequate, that the controller bears the burden of proving their adequacy, and that the exemption for an event not imputable to the controller is to be interpreted strictly (judgment of 14 December 2023, C-340/21, \emph{Natsionalna agentsia za prihodite}, ECLI:EU:C:2023:986). The external event decides neither fault nor exemption; the proof of measures does, and it falls on those who had to adopt them. This is the European model closest to the rule.

Directive (EU) 2024/2853 on liability for defective products treats software as a product and does not reduce the manufacturer's liability where a third party's act concurs (art.~13(1)), but it protects only natural persons, excludes damage to property and data used exclusively for professional purposes (arts.~5 and~6), and exempts the manufacturer that did not place the product on the market or put it into service (art.~11(1)(a)). An undertaking harmed during an evaluation, as in July 2026, would remain outside it. Where the delegator proves all suitable measures and no other head or person answers, the harm stays with the third party. The paper declares this gap; it is residual, because it arises only after every other action and every other responsible person has been examined.

\section{The delegation record as evidence}\label{sec:record}

If the excess is measured on the documented perimeter, the document matters. The delegation record keeps one row for each delegated capability, not one per agent: it is a register of acts of delegation, not a register of agents. Its minimal structure develops the seven-column matrix of \emph{Agentic AI} (capability, impact, risk on two rows, authorisation limit, human control, evidence, accountable person) \citep{fabiano2026agentic,fabiano2026acl}. The published version adds the verb on the five-verb scale with the threshold of irreversibility, the accesses, sub-delegation with depth and permissions, a stop mechanism with its gradability and route, the five functions of the accountable person (process owner, authoriser, custodian of the evidence, decision owner, person empowered to stop), an ethics row, and the dates of conferral, expiry and revocation \citep{fabiano2026adr}. A lint checks fifteen conditions as errors and five as warnings, including a mandatory threshold for irreversible verbs.

Each figure of excess maps to fields of the record: the verb, the limit and the control threshold, the accesses, the sub-delegation and the evidence of the chain. Compared with configurations and logs, the record allows the court to reconstruct the three perimeters and to locate the mode of exit. Its evidentiary value depends on five conditions: anteriority, integrity, attribution, correspondence with the configuration and identifiable validity in time. Where the agent processes personal data, the record is one of the documents through which the controller can discharge the burden described above.

Two limits must be stated. Documentation is not control: the record documents the control and preserves its proof, but it does not prevent anything. And its weight in court depends on the applicable head of liability, on the other evidence, and on the correspondence between what it declares and what happened. A record that declares a perimeter the configuration does not implement is, in the scenario of misconfiguration, evidence against the delegator.

\section{Validity and limitations}

The rule is formulated for civil liability. Criminal liability, the position of third parties relying on the agent's apparent authority, incident notification, the evaluation phase and the custody of evidence outside the agent's reach are outside the scope of this paper and are developed in the forthcoming book. The illustrative case is reconstructed from public sources, including an investigation whose scope was agreed with the organisation concerned; some facts relevant to the rule, such as when the estimate of unattainable tasks became known, are not established. Italian law serves as an example of national law; the transposition of the rule into other legal systems, including common law systems, is not attempted here. The record has not yet been tested in litigation.

\section{Conclusion}

Others have identified who should answer for an AI agent. This paper has proposed how that answer can be assessed: by imputing the excess to whoever conferred the delegation, measuring it on the perimeter that was documented and the control that was exercised, and letting the mode of exit decide what must be proved. The delegation record gives that measure a documentary form. The open question is no longer whether the organisation answers, but whether it can show what it authorised and how it governed it.

\section*{Availability}

The Agentic Delegation Records are published at \url{https://claraxis.eu/agentic-delegations/}, and the matrix implementation in the repository \url{https://codeberg.org/nicfab/agentic-control-layer} (EUPL-1.2). \copyright{} 2026 Nicola Fabiano. This preprint is distributed under the Creative Commons Attribution-NonCommercial-NoDerivatives 4.0 International licence (CC BY-NC-ND 4.0), \url{https://creativecommons.org/licenses/by-nc-nd/4.0/}.

\bibliographystyle{plainnat}
\bibliography{refs}

\end{document}